\documentclass{ws-ijmpa}
\usepackage[super,compress]{cite}
\usepackage{graphicx}
\begin{document}
\markboth{Victor Berezin, Vyacheslav Dokuchaev, Yury Eroshenko, Alexei Smirnov}{Least action principle and gravitational double layer}

%
\catchline{}{}{}{}{}
%

\title{Least action principle and gravitational double layer}

\author{Victor Berezin}
\address{Institute for Nuclear Research of the Russian Academy of Sciences, \\ 
Prospekt 60-letiya Oktyabrya 7a, Moscow, 117312 Russia \\
berezin@inr.ac.ru}

\author{Vyacheslav Dokuchaev}
\address{Institute for Nuclear Research of the Russian Academy of Sciences, \\ 
	Prospekt 60-letiya Oktyabrya 7a, Moscow, 117312 Russia \\
	dokuchaev@inr.ac.ru}

\author{Yury Eroshenko}
\address{Institute for Nuclear Research of the Russian Academy of Sciences, \\ 
	Prospekt 60-letiya Oktyabrya 7a, Moscow, 117312 Russia \\
	eroshenko@inr.ac.ru}

\author{Alexei Smirnov}
\address{Institute for Nuclear Research of the Russian Academy of Sciences, \\ 
	Prospekt 60-letiya Oktyabrya 7a, Moscow, 117312 Russia \\
	smirnov@inr.ac.ru}

\maketitle

\begin{history}
\received{\today}
\revised{Day Month Year}
\end{history}

\begin{abstract}
The higher derivative gravitational theories exhibit new phenomena absent in General Relativity. One of them is the possible formation of the so called double layer which is the pure gravitational phenomenon and can be interpreted, in a sense, as the gravitational shock wave. In this paper we show how some very important features of the double layer equations of motion can be extracted straight from the least action principle. 
\end{abstract}

\ccode{PACS numbers: 04.20.Fy, 04.20.Jb, 04.50.Kd, 04.60.Bc, 04.70.Bw.}


\section{Preamble}	

What is a double layer?

The very notion comes from the electrostatics, when one considers two oppositely charged very thin plates of infinite size. Each of them can be called the thin shell, and together they form the double layer. Note, that the plot of the charge distribution in this case resembles that of the approximation of the derivative of Dirac's delta-function.

In the Newton's theory of gravitation it is not possible to construct the double layer source, since the gravitational masses (charges) are all positive. Thus, two identical plates, when merging, will form the single thin shell with the mass doubled. The following note is in order here. The gravitational potential $\varphi$ obeys the Poisson equation $\Delta\varphi=4\pi G\rho$ ($\Delta$ is the Laplace operator, $\rho$ is the mass density, and $G$ is the Newton's constant), which is of the second order in derivatives. Let the gravitating mass is concentrated completely on the two-dimensional hypersurface $\Sigma_0$, and we denote by $n$ the coordinate normal to it ($\Sigma_0: n=0$). Then, the density $\rho=\sigma\delta(n)$. And the potential $\varphi$ can be written in the form
\begin{equation}
\varphi=\varphi_{(+)}\Theta(n)+\varphi_{(-)}(1-\Theta(n))
=(\varphi_{(+)}-\varphi_{(-)})\Theta(n) +\varphi_{(-)}= [\varphi]\Theta(n) +\varphi_{(-)},
\label{varphi}
\end{equation}
where $\varphi_{(\pm)}$ --- the potentials in $``+''(n>0)$ and $``-''(n<0)$ regions, and $\Theta(n)$ --- the Heaviside step function with the property $\Theta'(n)=\delta(n)$. Two different situations are possible. The first one is when the regions on the two sides are completely disjoined and must be considered separately as the spaces with the singular boundary $\Sigma_0$ each. In the second, we are dealing with the single space, divided into three parts, $``+''$, $``-''$ and $\Sigma_0$. Now $\Sigma_0$ is no merely the boundary. Physically, this means that the (test) particle are able (somehow) to penetrate through the singular hypersurface $\Sigma_0$. Mathematically, this means that one is allowed to differentiate the potential  (\ref{varphi}) at $n=0$. We have 
\begin{eqnarray}
\label{varphi1}
\varphi_{,n}&=&[\varphi]\delta(n)+[\varphi_{,n}]\Theta(n)
+\varphi_{(-),n},  \\
\varphi_{,nn}&=&[\varphi]\delta'(n)+2[\varphi_{,n}]\delta(n)
+[\varphi_{,nn}]\Theta(n)+\varphi_{(-),nn},
\label{varphi2}
\end{eqnarray}
where ``comma'' denotes the partial derivative. We see that there appears $\delta'(n)$, the plot of which resembles the double layer. But, their appears also the $\delta(n)$ in the expression for $\varphi_{,n}$, i.\,e., for the strength of the gravitational field. The latter provides the infinite point-like force acting on any test particle. It is attractive on one side and repulsive on another side of the singular hypersurface $\Sigma_0$, what is completely unphysical. This is why we have to put $[\varphi]|_{\Sigma_0}=0$. Therefore, there can be no double layer in Newtonian gravity.

Let us turn to General Relativity. The gravitational field is  now described by a second rank symmetric tensor $g_{\mu\nu}$ defining the space-time interval $ds^2=g_{\mu\nu}dx^\mu dx^\nu$, and replacing the single potential $\varphi$ in Newtonian gravity. The mass density $\rho$ is replaced by the matter energy-momentum tensor  $T_{\mu\nu}$. The single Poisson equation is replaced by the Einstein equations
\begin{equation}
R_{\mu\nu}-\frac{1}{2}g_{\mu\nu}R=\frac{8\pi G}{c^4}T_{\mu\nu}
\label{EE}
\end{equation}
(in what follows we will put $c=1$). Here $R_{\mu\nu}$ is the Ricci tensor, and $R$ is the curvature scalar, they are the convolutions of the Riemann curvature tensor $R^\mu_{\phantom{\mu}\nu\lambda\sigma}$: $R_{\mu\nu}=R^\lambda_{\phantom{\mu}\nu\lambda\sigma}$, $R=g^{\mu\nu}R_{\mu\nu}$. The left-hand-side of Einstein equations is of the second order in derivatives of the metric tensor (in the full analogy with Newton's theory). Now, let $\Sigma_0$ be some singular (3-dimensional) hypersurface, where the matter energy-momentum tensor is concentrated. Then 
\begin{equation}
T_{\mu\nu}=S_{\mu\nu}\delta(n)+[T_{\mu\nu}]\Theta(n)+T_{\mu\nu}^{(-)},
\label{T}
\end{equation}
where $n$ is a coordinate along the normal vector to $\Sigma_0$. Again, the whole space-time is divided into three parts, $(-)$, $(+)$ and $\Sigma_0$, and we are free to use quite different coordinate systems in the $(\pm)$-regions. But, we want to deal with the penetrable singular hypersurface $\Sigma_0$. So, if we assume that $[g_{\mu\nu}]|_{\Sigma_0}\neq0$, then we would obtain the $\delta$-function behaviour in the Christoffel symbols and, consequently, the $\delta^2$-behaviour in the Riemann curvature tensor, what is strictly forbidden by the conventional theory of distributions. Thus, in General Relativity there exists the purely mathematical reason to deal with the metric tensor, continuous on the singular hupersurface. Fortunately, by the coordinate transformations (different in different regions) one can always make the metric tensor continuous on any hypersurface $\Sigma$. Therefore, we are forced to work only with such special coordinate system (note, that this does not impose any restriction on the shape of a singular hypersurface). So, $[g_{\mu\nu}]|_{\Sigma_0}=0$. Then, the connections $\Gamma^\lambda_{\mu\nu}$ may undergo  a jump across $\Sigma_0$, and the Riemann curvature tensor $R^\mu_{\phantom{\mu}\nu\lambda\sigma}$ would have a $\delta$-function behaviour. Everything is self-consistent! But, still, there is no trace of a double layer!

All the previous consideration shows us that the gravitational double layer may appear only in the modified higher derivative gravity. One of these theories is the so called ``quadratic gravity''. The very name of it means that the gravitational Lagrangian contains  terms 
$R_{\mu\nu\lambda\sigma}R^{\mu\nu\lambda\sigma}$, $R_{\mu\nu}R^{\mu\nu}$ and $R^2$. Again, on the singular hypersurface $\Sigma_0$, defined by the equation $n=0$, we must impose $[g_{\mu\nu}]|_{\Sigma_0}=0$. But now, if $[g_{\mu\nu,n}]|_{\Sigma_0}\neq0$ is assumed, then one encounters with the $\delta$-function behaviour of the Riemann tensor and its convolutions, and, therefore, with the $\delta^2$-behaviour in the gravitational Lagrangian. What is forbidden by mathematics. Thus, we have to impose 
$[g_{\mu\nu,n}]|_{\Sigma_0}=0$. These are known as the Lichnerowicz conditions. One can write
\begin{equation}
g_{\mu\nu,nn}\!=\![g_{\mu\nu,nn}]\Theta(n)+\ldots, \;
g_{\mu\nu,nnn}\!=\![g_{\mu\nu,nn}]\delta(n)+\ldots, \;
g_{\mu\nu,nnnn}\!=\![g_{\mu\nu,nn}]\delta'(n)+\ldots \nonumber 
\label{gnnn}
\end{equation}
At last! We will call this the double layer.

In General Relativity,  we have the following classification. If there is a jump at $\Sigma_0$ in the energy-momentum tensor, which, of course, causes a jump in the space-time curvature, then we have the shock wave in the matter continuum accompanying by the gravitational shock wave. That is, one has a correspondence ``jump in curvature $\longleftrightarrow$ gravitational shock wave''.
If the matter is concentrated on $\Sigma_0$ ($\delta$-function behaviour), then it is called ``the thin shell''. In quadratic gravity a jump in curvature is reflected in appearance of the $\delta'$-function behaviour of the metric tensor, i.\,e., of the double layer. In the absence of the corresponding structure in the matter energy-momentum tensor we see that this is a pure gravitational phenomenon and it can be called ``the pure gravitational shock wave''.

\section{Least action principle and Israel-like equations}

The thin shell formalism in General Relativity was elaborated by Werner Israel more than 50 years ago \cite{Israel}. Its importance stems from the fact that Einstein equations are nonlinear, and obtaining exact solutions in the presence of matter becomes very difficult problem. The $\delta$-like distributions of the matter make this task easier. The exact solutions, found in such a way, provide us with necessary physical intuition.

Usually, Israel equations  are derived straight from the field equations. But, here we would like to demonstrate the power of the least action principle in doing the same. We find it necessary because this very technique we will be using for deriving the equations of motion for double layers.

We start with the Hilbert gravitational action $S_{\rm H}$ (throughout  the paper we will use the signature $(-,+,+,+)$ for the metric tensor)
\begin{equation}
S_{\rm H}=\frac{1}{16\pi G}\int\!\! R\sqrt{-g}\,d^4x.
\label{SH}
\end{equation}
The total action is the sum of the Hilbert action $S_{\rm H}$ and the action  for all other matter fields $S_{\rm matter}$. Since we are not interested in the overall boundary, there is no need in the Gibbons-Hawking boundary term. Thus, $S_{\rm tot}=S_{\rm H}+S_{\rm matter}$. By definition, the matter energy-momentum tensor $T^{\mu\nu}$ is
\begin{equation}
\delta S_{\rm matter} = \frac{1}{2}\int\!\! T^{\mu\nu}\delta g_{\mu\nu}\sqrt{-g}\,d^4x=
-\frac{1}{2}\int\!\! T_{\mu\nu}\delta g^{\mu\nu}\sqrt{-g}\,d^4x.
\label{enmomStot} 
\end{equation} 
The variation of the gravitational action reads 
\begin{equation}
\delta S_{\rm H} = \frac{1}{16\pi G}\int\left\{g^{\mu\nu}\delta R_{\mu\nu}+\right(R_{\mu\nu}-\frac{1}{2}g_{\mu\nu}R\left)\delta g^{\mu\nu}\right\}\sqrt{-g}\,d^4x.
\label{deltah} 
\end{equation} 
There exists the marvellous formula for the variation of Riemann curvature tensor, 
\begin{equation}
\delta R^\mu_{\phantom{\mu}\nu\lambda\sigma}=(\delta\Gamma^\mu_{\nu\sigma})_{;\lambda} -(\delta\Gamma^\mu_{\nu\lambda})_{;\sigma}.
\label{deltaRiemann} 
\end{equation}
(the semicolon denotes the covariant derivatives). With the use of the very well known relations $A^k_{;k}\sqrt{-g}=(A^\sigma\sqrt{-g})_{,\sigma}$ for any vector $A^\sigma$ and $g^{\mu\nu}_{\phantom{\mu\nu};\lambda}=0$ we get 
\begin{equation}
g^{\mu\nu}(\delta\Gamma^\lambda_{\mu\nu})_{;\lambda}= (g^{\mu\nu}(\delta\Gamma^\lambda_{\mu\nu}\sqrt{-g}))_{,\lambda}, \quad
g^{\mu\nu}(\delta\Gamma^\lambda_{\mu\lambda})_{;\nu}= (g^{\mu\nu}(\delta\Gamma^\lambda_{\mu\lambda}\sqrt{-g}))_{,\nu} 
\end{equation}
(note, that $\delta\Gamma$ is a tensor). Thus, the first term in the integrand is the full derivative and,  by virtue of the Stock's theorem, can be converted into the (boundary) surface integral. Also, it brings the contribution to the surface integral due to the $\delta$-function. The remaining part of the variation gives us the Einstein equations in the bulk. Nevertheless, we still need it because of the $\delta$-function hidden inside, that makes the contributions as well. The contribution from the full derivatives (from the bulk) equals
\begin{equation}
-\int_{\Sigma_0}\!g^{\mu\nu}[\delta\Gamma^\lambda_{\mu\nu}]\sqrt{-g}\,dS_\lambda+
\int_{\Sigma_0}\!g^{\mu\nu}[\delta\Gamma^\lambda_{\mu\lambda}]\sqrt{-g}\,dS_\nu.
\label{deltaGamma} 
\end{equation}
because inside the volume of integration we have some singular hypersurface $\Sigma_0$, where the matter energy-momentum tensor is concentrated, i.\,e., it has the $\delta$-function behaviour. Einstein equations translate this feature to the behaviour of curvature scalar $R$. It seems the easiest way would be to insert explicitly the corresponding part of $R$ into the integrand (\ref{SH}), and, after integration, to get the effective action, concentrated on $\Sigma_0$, which would serve as the starting point for the variational procedure. Let us do this step by step.   

The singular hypersurface $\Sigma_0$ divides the region of integration into 3 parts, (in)(=``$-$''),  
(out)(=``$+$'') and $\Sigma_0$. Let us introduce the so called Gauss normal coordinates associated with $\Sigma_0$ 
\begin{equation}
ds^2=\epsilon\,dn^2+\gamma_{ij}dx^idx^j,
\label{Gauss}
\end{equation}
where $x^i\in\Sigma_0$, $\epsilon=\pm1$ depending on whether $\Sigma_0$ is spacelike or timelike, and $n$ runs along its outward normal vector. Introduce also the extrinsic curvature tensor 
\begin{equation}
K_{ij}=-\frac{1}{2}\frac{\partial\gamma_{ij}}{\partial n}
\label{K}
\end{equation}
that describes the embedding of hypersurface $\Sigma_0$ into the 4-dimensional spacetime. 

In the Gauss normal coordinates this becomes 
\begin{equation}
-\int_{\Sigma_0}\!\left\{g^{\mu\nu}[\delta\Gamma^n_{\mu\nu}]-g^{\mu n}[\delta\Gamma^\nu_{\mu\nu}]\right\}\sqrt{|\gamma|}\,d^3x,
\label{deltaGamma2} 
\end{equation}
where $\gamma=\det||\gamma_{ij}||$. Very important note: one should realise that though $g_{nn}=\epsilon$ and $g_{ni}=0$ in the Gauss normal coordinates system, $\delta g_{nn}$ and $\delta g_{ni}$ are not necessarily zero, and the same is true for $\delta\Gamma^n_{nn}$,  $\delta\Gamma^n_{ni}$ and  $\delta\Gamma^j_{nn}$. Keeping this in mind, we get for the above expression
\begin{equation}
-\int_{\Sigma_0}\!\left\{\gamma^{ij}[\delta\Gamma^n_{ij}]-
\epsilon[\delta\Gamma^i_{ni}]\right\}\!\sqrt{|\gamma|}\,d^3x =-\int_{\Sigma_0}\!\!\epsilon\left\{\gamma^{ij}[\delta K_{ij}]+
[\delta K]\right\}\!\sqrt{|\gamma|}\,d^3x,
\label{deltaGamma3} 
\end{equation}
while all other terms appeared continuous and disappeared due to the $[\phantom{i}]$-operation.

To finish with the first term in the integrand of Eq.~(\ref{deltah}) we should find the contribution to the surface integral from its $\delta$-function part. Since
\begin{eqnarray} \label{Rnn}
R_{nn} &=& (K_{,n}-K^{lp}K_{lp})=[K]\delta(n) +\ldots, \\
R_{ij} &=& \epsilon(K_{ij,n}-K_{ij}K+2K_{ip}K^p_j)+\tilde R^{(3)}_{ij}
=\epsilon[K_{ij}]\delta(n) +\ldots,
\label{Rij}
\end{eqnarray}
we have
\begin{equation}
\epsilon\! \int_{\Sigma_0}\! ([\delta K] 
+\gamma^{ij}[\delta K_{ij}])\sqrt{|\gamma|}\,d^3x.
\label{surf} 
\end{equation}
Combining this with the result of Eq.~(\ref{deltaGamma3}) one gets zero. The terms with $\delta K_{ij}$ and $\delta K$ disappeared completely!

What is left? The following integral
\begin{equation}
\int (R_{\mu\nu}-\frac{1}{2}g_{\mu\nu}R)\delta g^{\mu\nu}\sqrt{-g}\,d^4x.
\label{deltah2} 
\end{equation}
from which one should extract the terms with $\delta$-function. Note that $(nn)$ and $(ni)$-components do not contain $\delta$-function at all. One finds immediately 
\begin{equation}
\delta S_{\rm H}(\Sigma_0)=\frac{\epsilon}{16\pi G}  \int_{\Sigma_0} (K_{ij}-\gamma_{ij}K)\delta\gamma^{ij}\sqrt{|\gamma|}\,d^3x.
\label{deltah3} 
\end{equation}
For the thin shell contribution one has (see Eq.~(\ref{enmomStot}))
\begin{eqnarray}
\delta S_{\rm matter}(\Sigma_0)&=& -\frac{1}{2}\int_{\Sigma_0}\!\! S_{\mu\nu}\delta g^{\mu\nu}\sqrt{|\gamma|}\,d^3x \nonumber \\
&=&  -\frac{1}{2}\int_{\Sigma_0}\!\! (S_{ij}\delta \gamma^{ij}
+S_{nn}\delta g^{nn}+2S_{ni}\delta g^{ni})\sqrt{|\gamma|}\,d^3x,
\label{deltaSmatter2}
\end{eqnarray}
The absence of the $\Sigma_0$ terms with $\delta g^{nn}$ and $\delta g^{ni}$ in $\delta S_{\rm H}(\Sigma_0)$ leads to the relations $S_{nn}=S_{ni}=0$. Finally, we arrive at the Israel equations $\epsilon([K_{ij}]-\gamma_{ij}[K])=8\pi G S_{ij}$.

\section{Least action principle and double layer in quadratic gravity}

We know already that the gravitational double layer may appear only in some modified higher derivative gravity. The most interesting representative of this type of the theories is the so called quadratic gravity with the Lagrangian
\begin{equation}
{\cal L}=\alpha_1  R_{\mu\nu\lambda\sigma}R^{\mu\nu\lambda\sigma}
+\alpha_2R_{\mu\nu}R^{\mu\nu}+\alpha_3R^2+\alpha_4R+\alpha_5\Lambda
\label{L} 
\end{equation} 
($\Lambda$ is the cosmological constant). The action is $S=\int\!{\cal L}\sqrt{-g}\,d^4x$. Again, inside the volume of integration there exists a singular hypersurface $\Sigma_0$, where the matter energy-momentum tensor may have a jump and/or the $\delta$-function behaviour. Also we know already that the metric tensor $g_{\mu\nu}$ is continuous on $\Sigma_0$, as well as the Christoffel symbols $\Gamma^\lambda_{\mu\nu}$ are. The derivatives of the connections may have jumps. They enter the Riemann curvature tensor $R^\mu_{\phantom{\mu}\nu\lambda\sigma}$ and its convolutions (Ricci tensor $R_{\mu\nu}$ and curvature scalar $R$) linearly, and it is these jumps that make possible the gravitational double layers to exist. 

The general mathematical formalism for a double layer in the quadratic gravity was elaborated by J.~M.~M.~Senovilla \cite{Senovilla}. We adopted it for the specific case of the spherically symmetric Weyl$+$Einstein gravity, when all the quadratic terms in the Lagrangian are combined to form the square of the Weyl tensor \cite{berdoker}. The simplification appeared so efficient that we are left on $\Sigma_0$ with only one component of 3-dimensional metric tensor and only one component of the extrinsic curvature tensor, namely, $\gamma_{00}$ and $K_{00}$. This allowed us to notice two very interesting and very important features of the matching conditions in the presence of double layer. First, they contain the completely arbitrary function and, second, the extrinsic curvature is zero. Both of these facts are of the same origin: the derivative of the $\delta$-function (i.\,e., $\delta'$) is not concentrated on $\Sigma_0$. But still, it is rather unusual. The aim of the present Section is to show how these features arise in the general case already at the level of the least action principle.

According to our strategy not to use explicitly the field equations in the bulk we will omit all terms (in the bulk) proportional to $\delta g^{\mu\nu}$ and $\delta g_{\mu\nu}$ whenever it is possible (i.\,e., if they do not contain $\delta$-function). Thus, the variation of that part of action, we really need, reads as follows
\begin{equation}
\delta S \quad \rightarrow \quad 
\int\!\delta{\cal L}\sqrt{-g}\,d^4x \quad\rightarrow
\nonumber
\end{equation} 
\begin{equation}
 \int \!(2\alpha_1R_\mu^{\phantom{\mu}\nu\lambda\sigma}\delta R^\mu_{\phantom{\mu}\nu\lambda\sigma}
+2\alpha_2 R^{\mu\nu}\delta R_{\mu\nu}
+2\alpha_3g^{\mu\nu}\delta R_{\mu\nu}
+\alpha_4g^{\mu\nu}\delta R_{\mu\nu})\sqrt{-g}\,d^4x.
\label{Sdouble}
\end{equation} 
Using the marvellous formula (\ref{deltaRiemann}) for $\delta R^\mu_{\nu\lambda\sigma}$ we are able to convert the above expression into
\begin{equation}
\int\!\!\left\{4\alpha_1R_\mu^{\phantom{\mu}\nu\lambda\sigma}
(\delta\Gamma^\mu_{\lambda\sigma})_{;\lambda}
\!+\!(2\alpha_2R^{\mu\nu}\!+\!2\alpha_3R\!+\!\alpha_4g^{\mu\nu})\!
\left((\delta\Gamma^\lambda_{\mu\nu})_{;\lambda}
\!-\!(\delta\Gamma^\lambda_{\mu\lambda})_{;\nu}\right)
\right\}\!\sqrt{-g}\,d^4x.
\nonumber 
\end{equation} 
Note that at this first stage we have no $\delta$-functions in the integrand, only jumps across the singular hypersurface $\Sigma_0$. And this makes difference from General Relativity.

The next step is to pick up the full derivatives and make use of the Stock's theorem. One gets
\begin{eqnarray}
\delta S \; \rightarrow \;
&&\int_{\Sigma_0}\!\left\{4\alpha_1R_\mu^{\phantom{\mu}\lambda\nu\sigma}
\delta\Gamma^\mu_{\lambda\sigma}+(2\alpha_2R^{\mu\lambda}+2\alpha_3g^{\mu\lambda}R
+\alpha_4g^{\mu\lambda})\delta\Gamma^\nu_{\mu\lambda}\right.\nonumber \\
&-&\left. (2\alpha_2R^{\mu\nu}+2\alpha_3g^{\mu\nu}R
+\alpha_4g^{\mu\nu})\delta\Gamma^\lambda_{\mu\lambda}\right\}\sqrt{-g}\,dS_\nu
\nonumber \\
&+&\int\!\left\{-4\alpha_1(R_\mu^{\phantom{\mu}\lambda\nu\sigma})_{;\nu}
\delta\Gamma^\mu_{\lambda\sigma}-(2\alpha_2R^{\mu\lambda}+\alpha_3R)_{;\nu}
\delta\Gamma^\nu_{\mu\lambda}\right.\nonumber \\
&+&\left. 2\,(\alpha_2R^{\mu\nu}+\alpha_3g^{\mu\nu}R)_{;\nu}
\delta\Gamma^\lambda_{\mu\lambda})\right\}\sqrt{-g}\,d^4x 
\end{eqnarray}
We see that there appeared (for the first time!) $\delta$-functions in the derivatives of the Riemann tensor and its convolutions. Remember now, that the surface integral eventually becomes
$\int_{\Sigma_0}\!(\phantom{x}) \; \rightarrow\; \int_{\Sigma_0}\![\phantom{x}]$.
Then, after introducing the Gauss normal coordinates,  $(n,x^i)$ and  $x^i\in\Sigma_0$, and removing the terms, continuous on $\Sigma_0$, we get 
\begin{eqnarray}
\delta S \; \rightarrow \;
&-&\int_{\Sigma_0}\!\left\{4\alpha_1[R_\mu^{\phantom{\mu}\lambda\nu\sigma}]
\delta\Gamma^\mu_{\lambda\sigma}
+(2\alpha_2[R^{\mu\lambda}]
+2\alpha_3g^{\mu\lambda}[R])\delta\Gamma^n_{\mu\lambda}\right.\nonumber\\
&-&\left. (2\alpha_2R^{nn}+2\alpha_3\epsilon[R])
\delta\Gamma^\lambda_{\mu\lambda}\right\}\sqrt{|\gamma|}\,d^3x
\nonumber \\
&+&\int\!\left\{-4\alpha_1(R_\mu^{\phantom{\mu}\lambda\nu\sigma})_{;\nu}
\delta\Gamma^\mu_{\lambda\sigma}-(2\alpha_2R^{\mu\lambda}+\alpha_3R)_{;\nu}
\delta\Gamma^\nu_{\mu\lambda}\right.\nonumber \\
&+&\left. 2\,(\alpha_2R^{\mu\nu}+\alpha_3g^{\mu\nu}R)_{;\nu}
\delta\Gamma^\lambda_{\mu\lambda})\right\}\sqrt{-g}\,d^4x 
\end{eqnarray}
Now, let us have look at the volume integral. The $\delta$-functions may appear only in the normal derivatives of jumps across $\Sigma_0$. And it is not very difficult to see, that the result of the integration will be equal exactly to the surface integral written above. What is the most important, the sign also will be the same (contrary to the situation in General Relativity)! The subsequent calculations are very tedious. So, we present here only some results, and the details will be published elsewhere in a separate paper. The final result for the surface integral is the following
\begin{equation}
-2\!\int_{\Sigma_0}\!\!\left\{\left((\!4\alpha_1\!
+\!\alpha_2)\gamma^{il}\gamma^{jp}\!
+\!(\alpha_2\!+\!4\alpha_3)\gamma^{ij}\gamma^{lp}\right)[K_{lp,n}]\delta K_{ij}\!+\!B_{km}\delta\gamma^{km}\right\}\!\!\sqrt{|\gamma|}\,d^3x,
\nonumber 
\end{equation}
where the exact expression for $B_{km}$ is not important here.

Let us first consider the exceptional case when the coefficient in front of $\delta K_{ij}$ is zero. Since $[K_{lp,n}]\neq0$ (the necessary condition for existing of the double layer),
\begin{equation}
4\alpha_1+\alpha_2=0, \quad \alpha2+4\alpha_3=0.
\label{alpha14}
\end{equation} 
This means that the double layer may appear, but it does not. With these conditions the quadratic gravity Lagrangian is nothing more but the famous Gauss-Bonnet term! In such a case the field equations are of the second order in derivatives of the metric tensor, not the fourth!

In the generic case, the variations $\delta K_{ij}$ and $\delta\gamma^{km}$ ($\delta\gamma_{km}$) are not independent on the singular hypersurface $\Sigma_0$, they are induced by the arbitrary variations of the trajectories in the $(\pm)$-regions. Thus, we can write
\begin{equation}
\delta K_{ij}=A_{ijkm}\delta\gamma^{km}.
\label{deltakij2}
\end{equation} 
And this is the way, how the completely arbitrary functions enter the equations of motion ($=$ matching conditions) of the double layer.

And the last thing. As already was mentioned, the main difference between $\delta'$-function and $\delta$-function is that the former is not concentrated on $\Sigma_0$. One of the consequence of this fact is the appearance of the arbitrary function. Another one is that the resulting tensorial equation will depend on whether we would raise the indices in variation of the metric tensor ($\delta\gamma^{ij}$ instead of $\delta g_{ij}$ or vice versa) before applying the Stock's theorem, or after. To remove such an inconsistency, we have to write 
\begin{equation}
K_{ij}|_{\Sigma_0}=0.
\label{kijsigma}
\end{equation} 
We postpone the detailed proof of this very stringent restriction on the shape of the singular hypersurface $\Sigma_0$ to the subsequent paper.

And the last note. In our approach we never explore the $\delta'$-function explicitly, only with $\delta$-function, in sequence.

\section*{Acknowledgments}

This work was supported in part by the Russian Foundation for Basic Research grant 18-52-15001-NCNIa.

\end{document}